\documentclass[journal]{IEEEtran}
\usepackage{amsmath,amsfonts}
\usepackage{algorithmic}
\usepackage{algorithm}
\usepackage{array}
\usepackage[caption=false,font=normalsize,labelfont=sf,textfont=sf]{subfig}
\usepackage{textcomp}
\usepackage{stfloats}
\usepackage{url}
\usepackage{verbatim}
\usepackage{graphicx}
\usepackage{cite}
\usepackage{siunitx}
\usepackage[T1]{fontenc}

\begin{document}

\title{Liquid-Crystal-Based Controllable Attenuators Operating in the 1--4~Terahertz Band}

\author{Aniela~Dunn,
Zhaopeng~Zhang,
Michael~D.~Horbury,
Eleanor~V.~Nuttall,
Yingjun~Han,
Mohammed~Salih,
Lianhe~Li,
Abigail~Bond,
Ehab~Saleh,
Russell~Harris,
Diego~Pardo,
Brian~N.~Ellison,
Andrew~D.~Burnett,
Helen~F.~Gleeson,
Alexander~Valavanis
\thanks{A.~Dunn and Andrew~D.~Burnett are with the School of Chemistry, University of Leeds, Leeds, LS2 9JT, UK}
\thanks{Z.~Zhang and Helen~F.~Gleeson are with the School of Physics and Astronomy, University of Leeds, Leeds, LS2 9JT, UK}
\thanks{M.~D.~Horbury, E.~V.~Nuttall, Y.~Han, M.~Salih, L.~Li, A.~Bond, and A.~Valavanis are with the School of Electrical and Electronic Engineering, University of Leeds, Leeds, LS2 9JT, UK}
\thanks{E.~Saleh and R.~Harris are with the School of Mechanical Engineering, University of Leeds, Leeds, LS2 9JT, UK}
\thanks{D.~Pardo, and B.~N.~Ellison, are with STFC Rutherford Appleton Laboratory Space Department, Didcot, OX11 0QX, UK}%
\thanks{Manuscript received May 23, 2023.}
}

\markboth{Dunn \MakeLowercase{\textit{et al.}}: Liquid-Crystal-Based Controllable Attenuators Operating in the 1--4~Terahertz Band}%
{Dunn \MakeLowercase{\textit{et al.}}: Liquid-Crystal-Based Controllable Attenuators Operating in the 1--4~Terahertz Band}


\maketitle

\begin{abstract}
Liquid-crystal devices (LCDs) offer a potential route toward adaptive optical components for use in the <\,2\,THz band of the electromagnetic spectrum.
We demonstrate LCDs using a commercially available material (E7), with unbiased birefringence values of 0.14--0.18 in the 0.3--4~THz band.
We exploit the linear dichroism of the material to modulate the emission from a 3.4\mbox{-}THz quantum cascade laser by up to 40\%, dependent upon both the liquid-crystal layer thickness and the bias voltage applied.
\end{abstract}

\begin{IEEEkeywords}
Liquid crystals, adaptive optics, variable attenuators, quantum-cascade lasers.
\end{IEEEkeywords}

\section{Introduction}
\label{Intro}
\IEEEPARstart{T}{he} development of terahertz (THz) technology has been motivated in part by the diverse range of potential applications including atmospheric and space research, biomedical and security imaging, and industrial inspection~\cite{Dhillon2017}.
However, THz adaptive optics (AO) technology is significantly less mature than in other spectral bands, and this potentially limits THz systems development within a range of practical scenarios.
AO systems employ components whose properties can be controlled dynamically to manipulate the wavefront of an optical field.
These enable, for example, automated image compensation for atmospheric turbulence~\cite{Jull2018}, laser power stabilization, beam-steering, dynamic beam focusing, polarization control~\cite{Milton2014}, and rapid single-pixel imaging~\cite{Stantchev2020}.
However, the relatively short wavelengths of THz radiation ($\sim$\SI{100}{\um}) introduce extremely challenging machining tolerances for the micro-electromechanical systems, deformable reflectors or micro-mirror arrays commonly used in millimeter-wave systems.
Furthermore, many materials used in infrared or visible optics are opaque at THz frequencies, and diffraction limits the use of microlens systems.

Nevertheless, previous studies into terahertz-frequency AO systems have both developed new techniques and adapted those used at other wavelengths.
Hybrid semiconductor--polymer structures~\cite{Hochberg2006}, interdigitated $p$--$n$ junctions~\cite{Ding2018}, hybrid split-ring resonators on both silicon and GaAs~\cite{Cong2018,Chen2006}, graphene metastructures~\cite{Degl'Innocenti2014}, and bismuth nanofilms~\cite{Song2021} have all been investigated at frequencies below \SI{2}{\THz}. 
These systems provide ultra-fast intensity modulation~\cite{Hochberg2006,Cong2018,Degl'Innocenti2014},
reflection cancellation~\cite{Ding2018},
and real-time control and manipulation of THz radiation~\cite{Chen2006,Song2021}, but they rely on complex manufacturing and lithography processes.

Liquid crystals (LCs), on the other hand, are a widely used technology in infrared and visible AO systems, based on relatively simple manufacturing techniques, and there is scope to extend their use to THz frequencies.
Several studies have already shown that liquid-crystal materials exhibit high birefringence at THz frequencies~\cite{Pan2003,Wilk2009,Vieweg2010,Chen2004,Yang2010,Park2012}, and devices have been manufactured for THz modulation~\cite{Savo2014,Schrekenhamer2013}, phase control~\cite{Wang2015,Yang2014,Yang2018,Ji2018,Li2020}, and frequency tuning~\cite{Shen2018,Yang2017} applications.
Aside from a few exceptions~\cite{Degl'Innocenti2014,Yan2012,Wen2014}, though, these studies were restricted to frequencies below \SI{2}{THz}. It therefore remains highly desirable to develop LC adaptive optics for use with either broadband THz sources, or high-powered narrowband quantum-cascade lasers (QCLs), which operate in the $\sim$2--5~THz range. 
Key potential applications include electrically-controlled attenuators, and shutters for remote sensing and imaging, or phase-shifters and waveplates for time-domain spectroscopy.

We have recently outlined preliminary findings that indicate that the commercially-available nematic LC material, E7, provides a large and controllable birefringence in the 2--5\,THz band, with scope for power modulation applications~\cite{dunn_development_2021, dunn_extracting_2022}.
However, at these frequencies, the wavelength of the radiation is comparable to typical LC film thickness, resulting in strong etalon effects.
In this work, we provide the first detailed thin-film THz time-domain spectroscopic analysis of the dielectric properties of E7 up to 4\,THz, accounting for the complex interplay with the device geometry.
We show that controllable attenuation can be achieved across the 2--5~THz band, based on the linear dichroism of the material.
We demonstrate power-modulation depths in excess of 40\% using a 3.4\mbox{-}THz quantum cascade laser source. 


\section{Fabrication of liquid-crystal devices}
\label{Fab}

\begin{figure}
    \centering
    \includegraphics{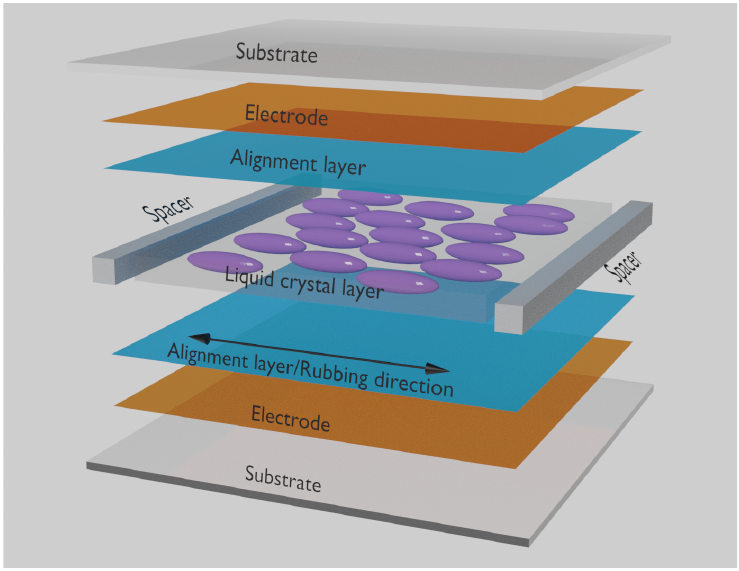}
    \caption{Showing the typical composition of the planar liquid-crystal cells used in these measurements. The cells consisted of two fused quartz windows in a parallel-plate arrangement, both with a conductive polymer [DMSO-doped poly(3,4-ethylene dioxythiphene) poly-4-styrene sulfonate] electrode layer and long-chain polyimide (SE-3510) alignment layer, separated from each other by a spacer to maintain the liquid-crystal layer thickness.}
    \label{fig:LCD}
\end{figure}

The devices fabricated for this study were based on an LC material layer enclosed within planar THz-transmissive cells, a schematic of which is shown in Fig.~\ref{fig:LCD}.
Although this work is focused on devices with controllable absorption, LC materials with high birefringence $\Delta{}n$ are generally desirable for AO components, as this enables large phase-retardation without the need for thick LC layers.
For these proof-of-concept devices, a commercially available nematic LC mixture, E7, was selected for this study, based on prior measurements of its large birefringence at visible wavelengths ($\Delta{}n=$ 0.21--0.26)~\cite{Chen2004,Yang2010} and from 0.2--2.0~\si{\THz} ($\Delta{}n=$ 0.13--0.15)~\cite{Pan2003,Wilk2009,Vieweg2010,Chen2004,Yang2010,Park2012}. Most materials used in the construction of conventional visible or infrared LCDs have poor transparency at THz frequencies.
As such, the devices developed in this study used alternative window and electrode materials.

\begin{table}[b]
\caption{List of LC devices fabricated in this work, indicating the nominal thickness of the LC layer and whether or not an electrode layer was included.}
\centering\begin{tabular}{l|r|l}
    \hline
    \hline
    Device ID & Nominal Thickness (\si{\um}) & Electrode \\
    \hline
    A & 320 & no  \\
    B & 100 & yes \\
    C & 13 & yes \\
    \hline
\end{tabular}
\label{tbl:devices}
\end{table}

Specifically, glass is typically used as a window material for visible LCDs. However, ionic impurities lead to ionic polarizability higher than that of materials such as silica glass~\cite{Naftaly2007}, making it a less desirable choice for THz applications. Fused quartz (silica) was used in the first paper describing an LC for THz applications~\cite{Chen2003}, and is widely used in many other applications. As such, fused quartz slides with a nominal thickness of \SI{1}{\mm} were chosen for the liquid crystal device (LCD) substrate.

Similarly, indium tin oxide layers, which are commonly used as electrodes in conventional LCDs, exhibit high THz absorption~\cite{Jewell2008,Lin2011}, and are practically opaque at thicknesses as small as tens of nanometres~\cite{Lin2011}.
Therefore, PEDOT:PSS, a conductive polymer [DMSO-doped poly(3,4-ethylene dioxythiphene) poly-4-styrene sulfonate]~\cite{Du2016,Du2016a,Sasaki2017}, was chosen as the electrode layer material, as it provides a transmittance of up to 83.5\% at \SI{1.22}{\THz}~\cite{Yang2014}, decreasing to approximately 70\% at \SI{2.5}{THz}~\cite{Yan2015}. More recent studies have shown that PEDOT:PSS has a relatively constant transmittance from 1--\SI{6}{\THz}~\cite{Dutin2019}, and a low-frequency conductivity comparable to a sputtered indium tin oxide thin film~\cite{Yang2014}.  
The PEDOT:PSS layer was spin-coated onto the fused quartz substrate, resulting in a thickness on the order of 10s of nanometres. 

Three LC devices, labeled A--C (see Table~\ref{tbl:devices}), were fabricated for this study, using the materials described above. Device A was manufactured using a thick LC layer (\SI{320}{\um}) and without an electrode layer to allow  accurate characterization of THz spectral parameters of the bulk LC material from 0.3--4.0~\si{\THz}. Devices B and C were manufactured using thinner LC layers (\SI{100}{\um} and \SI{13}{\um}, respectively) and with electrode layers, to enable characterization of the effect of an applied LC bias voltage on THz transmission.

The thicker devices (A \& B) present a challenge for fabrication, as conventional surface-alignment techniques cannot readily achieve good LC mono-domain alignment at this scale. Therefore, to promote alignment, a 4\% long-chain polyimide (SE-3510) layer was spin-coated onto the electrode layer and then rubbed to orientate the polymer, confining the director along the rubbing direction and parallel to the substrate surface.

The LCDs were assembled in a parallel-plate arrangement: two fused quartz/electrode layer (where used)/alignment layer substrates were sandwiched together, with the alignment layers facing, and separated from each other using a spacer to maintain the LC layer thickness.
The E7 material was then capillary filled into the empty cell.
Wires were connected to the two electrode layers of LCDs (where used) using both indium soldering and UV cured glue to provide a robust electrical connection. Reference samples of the quartz slide and polyimide layers were used to determine the refractive indices and absorption coefficients of these materials to aid in building a transfer function for the LCD. 

\section{Experimental characterization}
\label{Exp}

\begin{figure*}[t]
\centering
\subfloat[]{\includegraphics[width=\columnwidth]{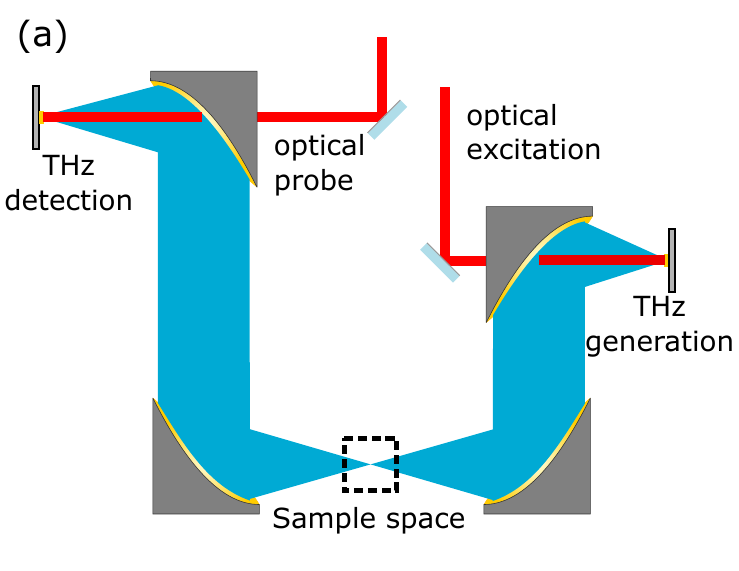}}
\hfil
\subfloat[]{\includegraphics[width=\columnwidth]{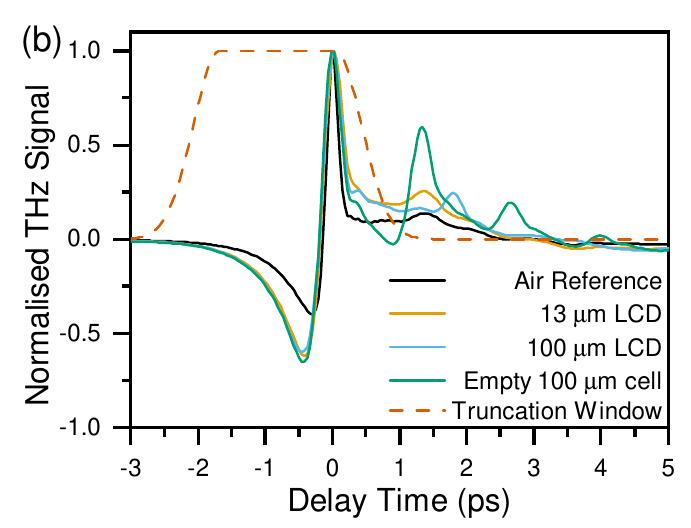}}
\caption{(a) Schematic diagram showing the system used to perform broadband terahertz spectroscopy measurements on the liquid-crystal devices. Time-domain traces acquired using this system are shown in (b), displaying a purged-air terahertz reference (black line), the terahertz response through the \SI{13}{\um} and \SI{100}{\um} liquid-crystal devices (yellow and blue lines respectively), the terahertz response through an empty liquid-crystal cell with a \SI{100}{\um} air-gap instead of a liquid-crystal layer (green line), and the truncation window applied to the data to remove the effects of reflections in the time-domain (dashed orange line).}
\label{fig:exp_bb}
\end{figure*}

Terahertz (THz) transmission measurements were performed on each of the LCDs using a broadband THz time-domain spectroscopy (TDS) system, which provides a free-space bandwidth of 0.3--\SI{8}{\THz}~\cite{Bacon2020} using an ultrafast Ti/sapphire Vitara--HP (Coherent) laser. 
However, a phonon mode in fused quartz causes an absorption at $\sim$5~THz, which effectively limits the measurement bandwidth in these samples to approximately \SI{4}{THz}.
The 800\mbox{-}nm output (\SI{20}{fs} pulse width, \SI{80}{MHz} repetition rate) from the laser was separated into two beams, where 90\% was used to generate horizontally-polarized THz radiation from a low-temperature-grown GaAs (LT-GaAs) bow-tie-shaped photoconductive emitter on a 2\mbox{-}mm-thick quartz substrate~\cite{Bacon2016}, with an electrode spacing of \SI{100}{\um}. This was biased using a 7\mbox{-}kHz AC square wave.
This modulation signal was also used as a reference frequency for lock-in detection.
The THz radiation generated from the emitter was collected in a backwards geometry (\textit{i.e.}, from the same surface of the emitter that was excited by the laser), thus avoiding absorption and dispersion in the undoped LT-GaAs and quartz substrate, and improving the high-frequency components of the THz spectrum.

The emitted THz radiation was collected and focused onto the sample using a pair of off-axis parabolic mirrors.
A second pair of off-axis parabolic mirrors was used to collect the THz radiation transmitted through the sample and focus it onto an LT-GaAs photoconductive detector, identical to the photoconductive emitter, alongside the remaining 10\% of the laser as an optical probe. A diagram of this is shown in Fig.~\ref{fig:exp_bb}(a).
The transient current generated in the photoconductive detector by the interaction of the THz and 800\mbox{-}nm beams was amplified using a low-noise pre-amplifier at 50~nA/V and measured using a lock-in amplifier.
The system was located within an enclosure that allows the atmosphere to be purged with dry air to a relative humidity of $<$2\%, minimizing attenuation by water absorption lines in the THz spectra.
The LCDs were rotated to orientate the LC director to be either perpendicular or parallel to the polarization direction of the THz radiation, to measure the ordinary and extraordinary refractive indices and absorption coefficients respectively. 

\subsection{Material properties at THz frequencies}
\label{MatProp}

\begin{figure*}[htbp]
\centering\includegraphics{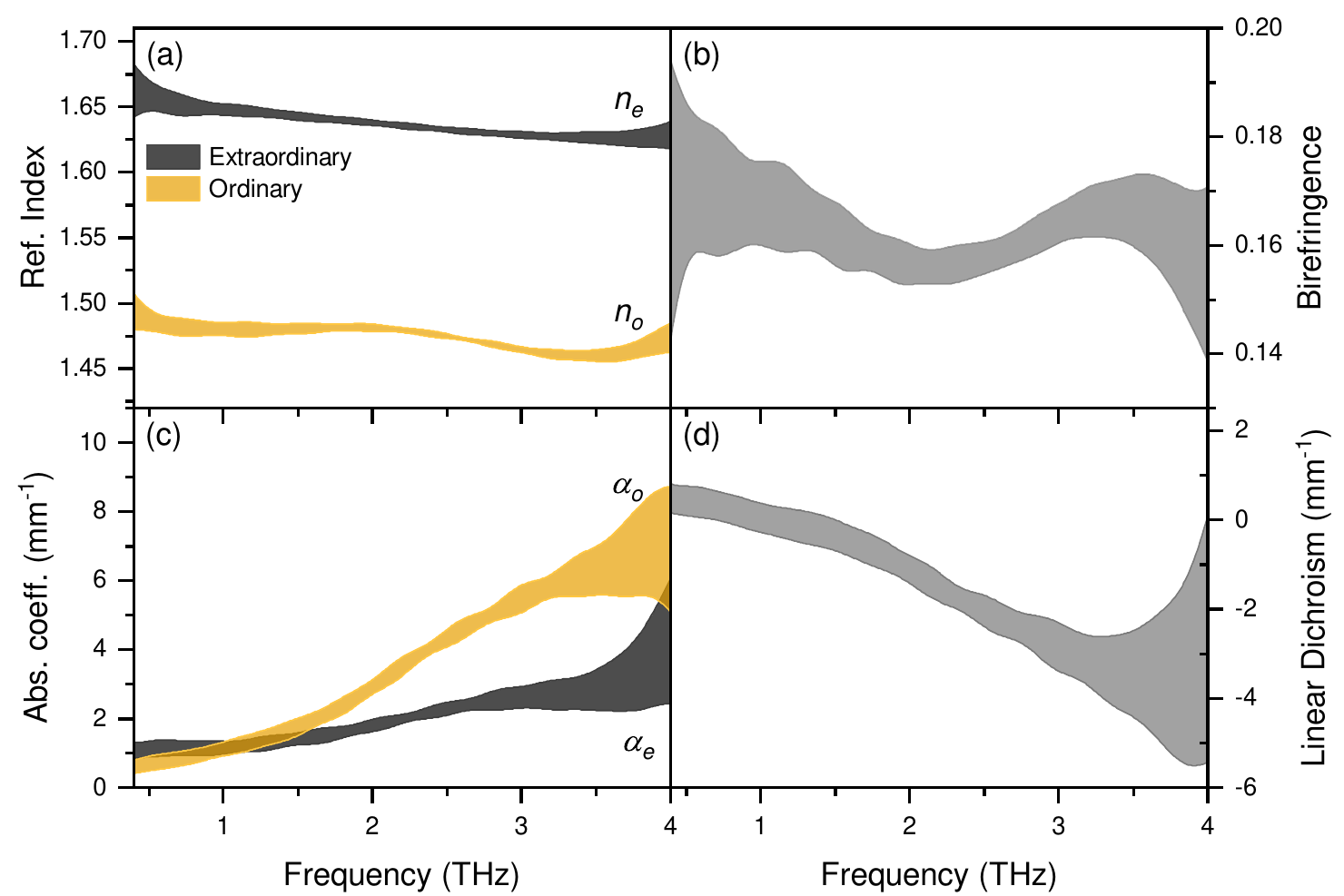}
\caption{\label{fig:E7_props}(a) The ordinary and extraordinary refractive indices of a planar sample of the LC material, E7, extracted from terahertz time-domain measurements of a LCD with a \SI{320}{\um} LC layer thickness (Device A). The calculated birefringence ($\Delta n = n_e - n_o$) is shown in (b). The corresponding absorption coefficients of the ordinary and extraordinary axes are shown in (c) and the linear dichroism ($\alpha_o - \alpha_e$) is shown in (d). The shaded areas indicate the error on the measurements based on the variability across repeat measurements.}
\end{figure*}

Device A (320\mbox{-}\si{\um} LC layer) was designed to characterize the E7 material at extended THz frequencies, as thicker material layers provide a stronger THz response, increasing the signal-to-noise ratio of the obtained spectral parameters, while omitting electrode layers reduces the complexity of the spectral analysis.
The complex refractive indices of the LC material were extracted numerically from the THz-TDS signal by the fitting of a transfer function using the data processing tool Nelly~\cite{Tayvah2021,Nelly2021}.

Examples of the time-domain signals that were acquired from the broadband THz spectroscopy can be seen in Fig.~\ref{fig:exp_bb} (b), which shows a purged-air reference (black line), the THz transmission measured through the 13\mbox{-}\si{\um} and 100\mbox{-}\si{\um} liquid-crystal devices (yellow and blue lines respectively) and the THz transmission though an empty liquid-crystal cell with a 100\mbox{-}\si{\um} air-gap instead of a liquid-crystal layer (green line). 

In the time-domain traces, shown in Fig.~\ref{fig:exp_bb}, reflections of the THz radiation occurring from each interface within the sample are easily identifiable. The fused quartz window materials act as an etalon, reflecting approximately 5\% of the THz signal at each LC/electrode/quartz interface and 10\% of the THz signal at the quartz/air interfaces.
The LC/electrode/quartz reflections are seen as small peaks in the time-domain signal at 1.3~ps (and subsequent reflections at 2.3~ps and 3.9~ps) and 0.15~ps after the main THz signal for the 100\mbox{-}\si{\um} and 13\mbox{-}\si{\um} LCDs respectively. This effect is exaggerated for the empty 100\mbox{-}\si{\um} LC cell, as the Fresnel coefficients for the reflections of the THz radiation between the two fused quartz windows is increased, owing to the lower refractive index of air compared to the LC material E7. The quartz/air reflections are seen approximately 13~ps after the main signal for both samples (not shown). 

Typically, time-domain data are windowed to remove reflections from the sample/air interfaces, minimizing the oscillations seen in the frequency domain. However, for a multilayered structure, especially one that contains thin layers such as these LCDs, removal of reflections can be challenging. We also chose to use a free-space air measurement for the reference.
However, as noted by others~\cite{Tayvah2021,Kolbel2022,Li2021}, this method can lead to both uncertainty in phase unwrapping during the transfer function fitting, along with an improper cancellation of etalons when fitting, particularly for high bandwidth measurements. 

As such, the data was truncated and zero padded to minimize the influence of etalons caused by reflections between layers in the extracted complex permittivity of the LC layer. An example of a truncation window is shown in Fig.~\ref{fig:exp_bb}(b) as a dashed orange line, which removed the etalon effects by restricting the time-domain signal to approximately \SI{1}{ps} after the main THz peak. It is important to note that the 320\mbox{-}\si{\um} LCD (Device~A) did not contain electrode layers and as such did not exhibit these etalon effects, and only the quartz/air reflections approximately \SI{13}{ps} after the main THz peak needed to be accounted for. 

Initial reference measurements of fused quartz and polyimide-on-quartz were used to determine their complex refractive index. This information was then used to define a suitable transfer function for Device~A. 

The thicknesses of each layer were then determined by means of a total variance analysis~\cite{Pupeza2007} of the transfer function fitting of the unwindowed time-domain data trace (70\mbox{-}\si{\ps} scan range) \textit{i.e.}, containing all interface reflections. Here the measurement of the cell was used as the sample and a purged measurement without the cell was used as a reference. From this, the LC thickness was determined to be 309~$\pm$~\SI{1}{\um}, compared to the nominal LC thickness of \SI{320}{\um}.

As can be seen from Fig.~\ref{fig:E7_props}(a), the ordinary and extraordinary refractive indices are distinct from one another, suggesting reasonable mono-domain alignment of the LC layer, even in this thicker device. The magnitudes of these refractive indices are slightly lower than those previously published~\cite{Yang2010,Vieweg2010,Park2012}, outside three standard variations, however, the authors note there is already a slight variation in the literature values depending on the sample thickness and data extraction method used. While the values measured in this work appear slightly lower than previously published, it is most likely because of the use of total variance analysis used in this work, allowing the thickness of the LCD layers to be obtained \textit{in situ}, which the other published works do not include. It is observed that both $n_o$ and $n_e$ remain relatively flat above 2~THz which is consistent with the behavior observed below 2~THz~\cite{Yang2010,Vieweg2010,Park2012}.

A birefringence of $\Delta{}n = $ 0.14-0.18 was observed between \SI{0.3}{THz} and \SI{4.0}{THz} by comparing the real extraordinary and ordinary refractive indices of the E7 LC material. These values are comparable to those previously measured for E7 below 2~THz~\cite{Chen2004,Yang2010,Park2012}, and are shown in Figure~\ref{fig:E7_props}(b). The absorption coefficients for E7 are shown in Figure~\ref{fig:E7_props}(c) and display linear dichroism, with the extraordinary axis displaying a lower absorption coefficient than the ordinary axis, again consistent with previous measurements at lower THz frequencies. This dichroic effect increases with THz frequency, shown in Figure~\ref{fig:E7_props}(d). Errors on these measurements were estimated by combining an analysis of the influence of time-domain truncation on the extracted complex permitivity, with the uncertainty of the thickness of the LC layer and repeat measurements.   

\begin{figure*}[tb]
\centering\includegraphics{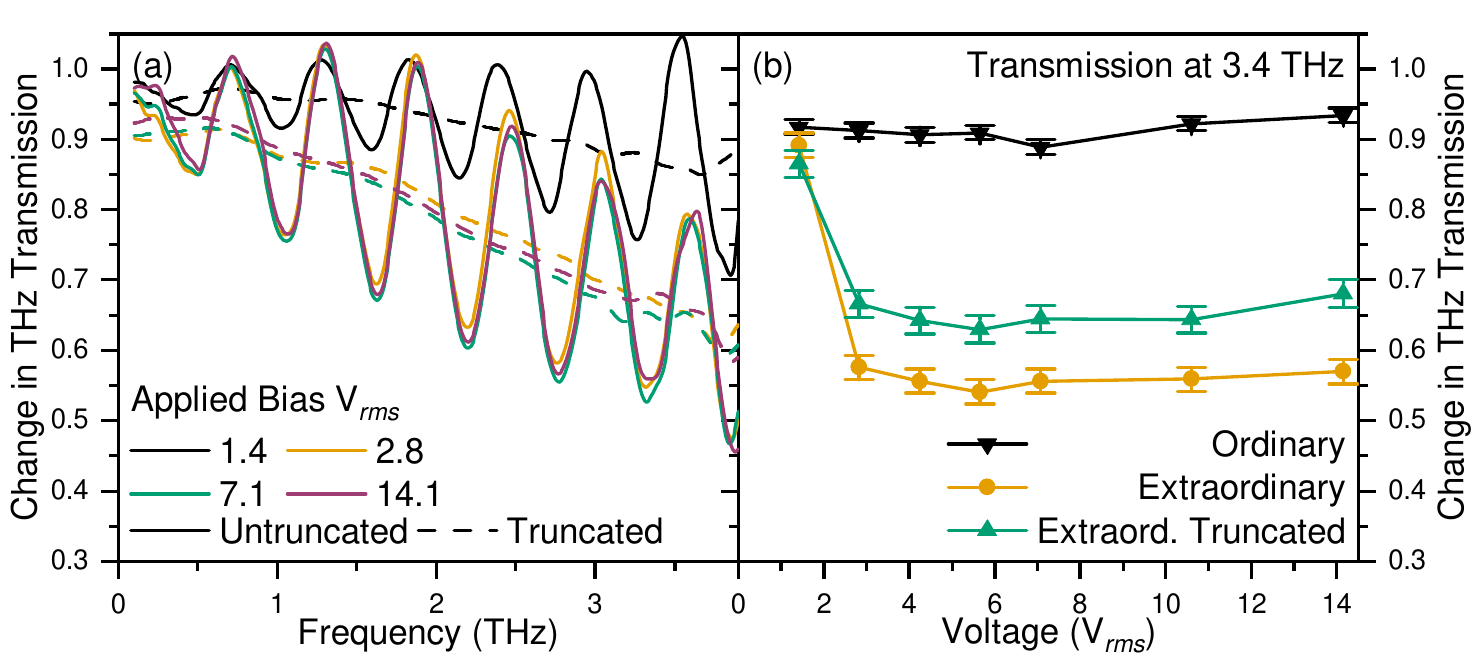}
\caption{\label{fig:transmission}(a) Terahertz transmission through the extraordinary axis of the \SI{100}{\um} LCD (Device B) for untruncated (solid lines) and truncated (dashed lines) data at different LC bias voltages. The THz transmission through the \SI{100}{\um} LCD at 3.4~THz is shown in (b), for both the ordinary (black down-triangles) and extraordinary (untruncated: yellow circles; truncated: green up-triangles) axes.}
\end{figure*}

Devices B and C, with 100\mbox{-}\si{\um} and 13\mbox{-}\si{\um} LC layer thicknesses respectively, were characterized to analyze the effect of applying a 5\mbox{-}kHz sinusoidal voltage from a signal generator to wires connected to the electrode layers.
Initially, the devices were investigated using Polarized Optical Microscopy (POM), which confirmed good quality alignment, with the PEDOT:PSS electrodes working well over the device area, and the E7 material responding uniformly to an applied electric field. THz-TDS measurements were performed for both LC axes at a range of bias voltage amplitudes ($V_{\text{rms}}$). To demonstrate the effect of biasing the LC layer, Fig.~\ref{fig:transmission} shows the relative change in THz transmission of Device B (\SI{100}{\um}) upon biasing for both the extraordinary and ordinary axes, relative to the unbiased device in the respective axis. 

All data presented in Fig.~\ref{fig:transmission}(a) have been windowed to remove the etalons associated with the quartz/air interface reflection approximately \SI{13}{ps} after the main peak, as these reflections will be unaffected by the bias voltage applied to the LC layer. However, the etalons associated with the LC/electrode/quartz interface are dependent on the refractive index of the LC layer which will vary as a bias is applied across the device. As such, analysis was performed for both the untruncated (including LC/electrode/quartz etalon effects) and truncated (etalon effects removed) data.

Figure~\ref{fig:transmission}(a) shows the change in the extraordinary transmission through the 100\mbox{-}\si{\um} LCD (Device B) as a function of applied bias for both the untruncated (solid lines) and truncated (dashed lines). As the bias is ramped up to  2.8\,V$_{\text{rms}}$ the LC layer begins to align with a clear change in the THz transmission through the device. After this point the LC layer has almost completely aligned with the field direction and further increasing the bias has a negligible effect on the THz transmission of the device. 
Strong oscillations in the untruncated data (solid lines) presented in Fig.~\ref{fig:transmission}(a) are attributed to the etalon effects of LC/electrode/quartz interface, with the etalon peaks' centers, widths, and amplitudes all changing as the effective birefringence of the LC changes (which in turn causes a change in the Fresnel Coefficients at the interface). By truncating the data before analysis, it is possible to separate the etalon effects from the response of the LC layer itself. 
To compare with the narrowband measurements discussed later in this work, the amplitude of the calculated change in THz transmission at 3.4~THz is shown in Fig.~\ref{fig:transmission}(b) for the ordinary axis (black down-triangles), and the extraordinary axis, both untruncated (yellow circles) and truncated (green up-triangles). It is important to note that the data in Fig.~\ref{fig:transmission}(b) are normalized using the transmission through the unbiased LCD along their respective axes \textit{i.e.}, the (extra)ordinary data is normalized using the transmission through an unbiased reference along the (extra)ordinary axis. Minimal change is seen for the ordinary THz transmission as a function of increasing bias voltage, whereas a large change is seen in the transmission on the extraordinary axis as the voltage bias is applied, stabilizing above 2.8~V$_{\text{rms}}$ for both the truncated and untruncated data. Errors were estimated based on the variability across repeat measurements.  

The frequency dependence of the change in THz transmission for the truncated data follows that of the linear dichroism shown in Fig.~\ref{fig:E7_props}(d). As such it is reasonable to assume that the change in absorption coefficient of the LC material under applied bias is responsible for the change in THz transmission through the LCD, rather than any birefringent effects.
Indeed, the 100\mbox{-}\si{\um} LC film is too thin compared with the wavelength range of the incident radiation (1000--\SI{75}{\um}) for birefringence to introduce significant phase-retardation between the ordinary and extraordinary components.
At \SI{3.4}{THz}, the linear dichroism is ($-3.4\pm0.8$)\,\si{\per\mm}, which would give $\sim$30\% lower absorption along the ordinary axis than the extraordinary axis of the 100\mbox{-}\si{\um} LCD.
This is comparable to the amplitudes of the truncated extraordinary axis data shown in Fig.~\ref{fig:transmission}(b). For the untruncated data, a further 10\% decrease is seen in the THz transmission as a result of biasing the LC layer, highlighting that interference from the change in Fresenel reflection coefficients at the LC/electrode/quartz interface also plays a substantial part in the modulation of THz radiation through the biased LC layer.

Although the 13\mbox{-}\si{\um} LCD was also investigated in this manner, the response was small, compared with the uncertainty of the measurement, meaning that no THz transmission data could be  determined reliably from the THz-TDS measurements.

\subsection{Terahertz modulation analysis}
\label{THzMod}

\begin{figure}
    \centering
    \includegraphics{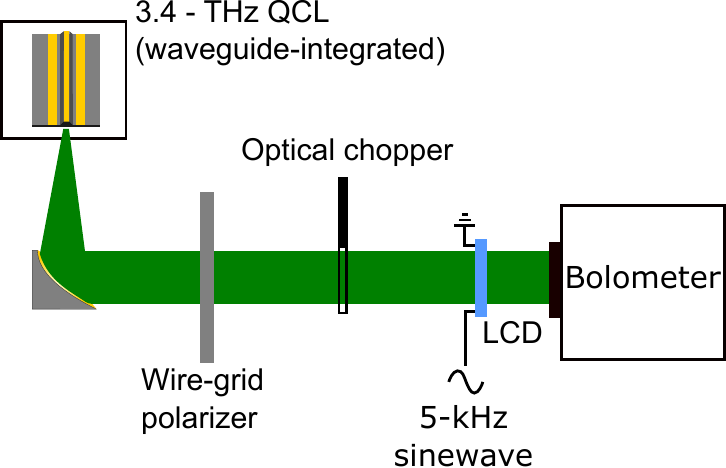}
    \caption{Schematic diagram of the experimental setup used for narrowband characterization of the liquid-crystal devices. The output from a  3.4\mbox{-}THz QCL, mounted in a helium-cooled cryostat, was collected and collimated with an off-axis parabolic mirror and passed through a wire-grid polarizer to ensure linearity of the polarized terahertz beam. This linearly polarized beam was optically chopped at \SI{167}{Hz} and subsequently passed through the liquid-crystal devices that were biased with a \SI{5}{kHz} sinusoidal voltage. The transmitted terahertz radiation was detected by a helium-cooled Si-bolometer, which was referenced at the chopper frequency.}
    \label{fig:exp_nb}
\end{figure}

\begin{figure*}[tb]
\centering\includegraphics{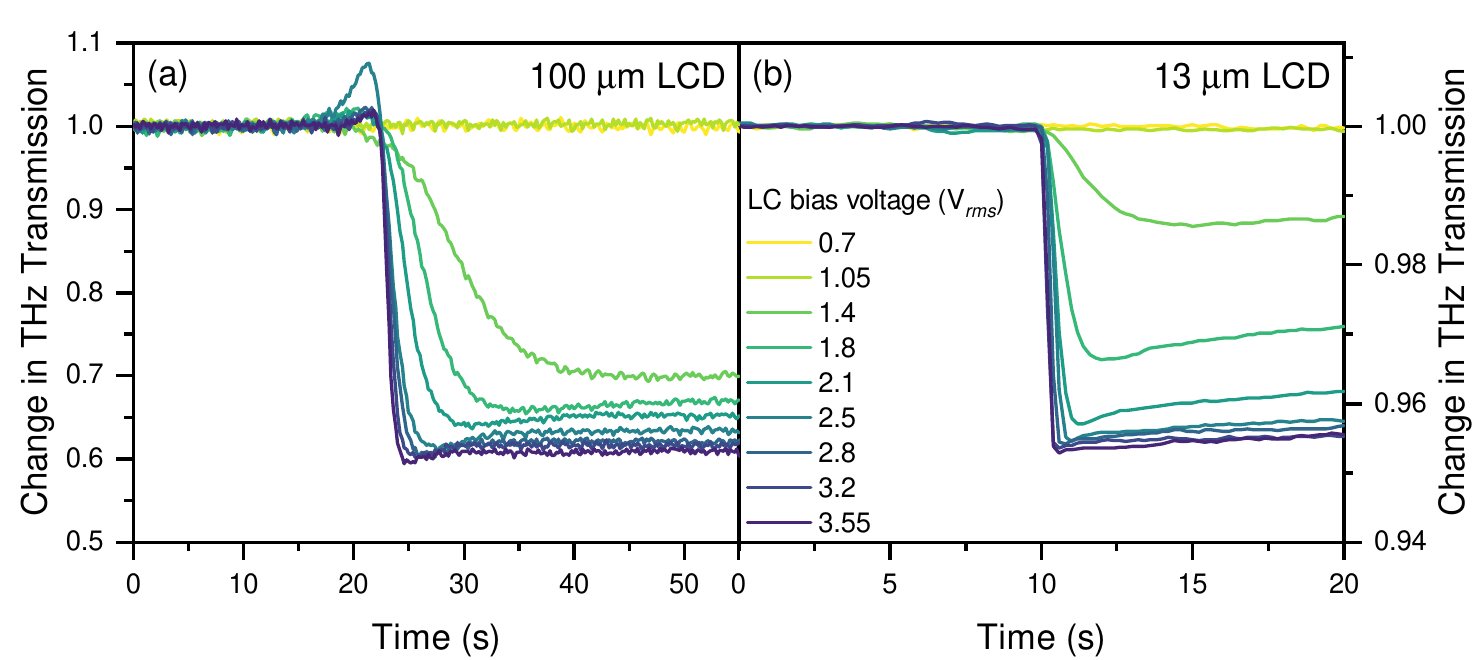}
\caption{\label{mod_depth}THz transmission through the extraordinary axis of (a)~the 100\mbox{-}\si{\um} LCD (Device~B) and (b)~the 13\mbox{-}\si{\um} LCD (Device~C) as a function of the $V_{\text{rms}}$ voltage applied.
The liquid crystal director was orientated parallel to the polarization direction of the collimated THz beam.}
\end{figure*}

A 3.4\mbox{-}THz waveguide-integrated QCL, described in Ref.~\cite{Ellison2019}, was selected as a narrowband characterization source, with a total $\sim$15\mbox{-}GHz tunability across several emission modes, controlled by varying the bias current applied to the QCL. The total QCL bandwidth is much narrower than the free-spectral range of the etalons in the LCD, allowing us to still treat this as a narrowband measurement technique. To determine the effect of the LCD as a THz modulator, the continuous-wave power output from the QCL was collected and collimated using an off-axis parabolic mirror, before being passed through a wire-grid polarizer (Microtech Instruments) to ensure a linearly polarized THz beam. This linearly polarized beam was optically chopped at 167~Hz and passed through the liquid crystal devices perpendicular to their surfaces, before being detected using a helium-cooled Si-bolometer (QMC Instruments). The bolometer signal was monitored using a lock-in amplifier, which was referenced to the chopper frequency. The LCDs were again biased with a 5\mbox{-}kHz sinusoidal voltage and could be rotated freely within the THz beam. This is shown in Fig.~\ref{fig:exp_nb}.

As with the THz-TDS measurements, it was observed that the LCD strongly attenuated the THz radiation, even without a bias applied to the liquid crystal layer.
Approximately 99.7\% and 99.83\% of the incident THz signal was lost through the 100\mbox{-}\si{\um} and 13\mbox{-}\si{\um} LCDs respectively, as a result of both reflection and absorption of the THz radiation. The greatest contributor to this loss in THz transmission through the LCDs is the fused quartz window material; a single 1\mbox{-}mm-thick fused quartz window was seen to attenuate the THz signal in THz-TDS measurements by 34\%, with approximately 11\% reflection loss at each air/quartz interface, and the remainder lost through absorption within the quartz window itself. As QCLs are relatively powerful sources of THz radiation, the transmitted THz radiation through the LCDs still gave a large enough transmitted intensity to easily allow the effect of biasing the LC layer to be observed. 

The bolometer signal was monitored as a function of time, as the bias across the LCD was varied.
Measurements were taken with the LC director orientated both parallel (ordinary axis) and perpendicular (extraordinary axis) to the polarization direction of the collimated THz beam, and also as a function of the incident THz power, \textit{i.e.}, the bias applied to the QCL device.

Figure~\ref{mod_depth} shows the THz transmission through (a)~the 100\mbox{-}\si{\um} LCD (Device~B) and the (b)~13\mbox{-}\si{\um} LCD (Device~C) for the situation in which the LC director was orientated parallel to the polarization of the collimated THz beam \textit{i.e.}, the extraordinary axis.
Below a certain LC bias-voltage threshold ($\approx$1~V$_{\text{rms}}$), no change is observed in the transmitted THz radiation through both of the LCDs.
However, above this threshold, the THz transmission through both devices decreases for increasing bias voltage.
A maximum change in the THz transmission \textit{i.e.}, a modulation of the THz signal, of 40\% and 4.5\% is seen for the 100\mbox{-}\si{\um} and 13\mbox{-}\si{\um} LCDs respectively. The measured THz transmission values shown in Fig.~\ref{mod_depth}(a) for the the \SI{100}{\um} LCD are consistent with the values determined from THz-TDS in Fig.~\ref{fig:transmission}. 

A slight transient increase in the transmission can be seen at some biases for the the \SI{100}{\um} LCD. This is likely explained by the etalon shifts that occur as the LC material responds to the applied bias.  It can be seen from Fig.~\ref{fig:transmission}(a) that these etalon oscillations can lead to transmission greater than the unbiased value. However, the observation of such an effect could be complicated by the switching speed within the LC layer (which is inversely proportional to layer thickness), the result of multiple domains forming within the LC layer, and/or the sampling speed of the measurements themselves, which could reduce the appearance of an effect that occurred more quickly than the smallest time step.

Using the linear dichroism values calculated for the 100\mbox{-}\si{\um} LCD, it is possible to estimate the maximum attenuation of the THz radiation through the the 13\mbox{-}\si{\um} LCD at 3.4~THz. This gives a value of (4~$\pm$~1)\%, which is comparable with the values measured through experiment and shown in Fig.~\ref{mod_depth}(b). 
When the LC director was perpendicular to the polarization direction of the THz radiation (\textit{i.e.}, the ordinary axis) there was no change in the THz transmission through the device for increasing bias voltage, consistent with the THz transmission data shown in Fig.~\ref{fig:transmission}(b) for the ordinary axis. 

The speed of this modulation depends on both the bias voltage applied to, and the thickness of, the LC layer with minimum fall times of (0.9~$\pm$~0.1)\,s above a 2.5\,V$_{\text{rms}}$ bias, and (5.0~$\pm$~0.2)\,s at a 14\,V$_{\text{rms}}$ bias for the 13\mbox{-}\si{\um} and 100\mbox{-}\si{\um} LCDs respectively. 
Rise times (data not shown) were also observed to increase with the modulation depth, on the order of a few seconds for the 13\mbox{-}\si{\um} device and upwards of 400\,s for the 100\mbox{-}\si{\um} device at maximum biasing conditions (7\,V$_{rms}$ and 14\,V$_{\text{rms}}$ respectively).
The modulation depths and speeds were found to be independent of the THz source power and as such the modulation is considered to be independent of frequency over the $\sim$15\,GHz tuning range of the QCL, as expected from the relatively flat absorption coefficients at this range of frequencies.

\section{Conclusions and further work}
The optical properties of the commercially available liquid crystal material E7 have been characterized up to \SI{4.0}{\THz}, with a birefringence of 0.14--0.18 for an unbiased device.
It was determined that the change in THz transmission through the LC layer could be explained by a combination of linear dichroism and etalon interference from the LC/electrode/quartz interface, of which both effects have significant influence on the LCD acting as a THz modulator. 
These liquid crystal devices have also been used to modulate the output of a THz QCL to depths of up to 40\%, determined by the LC layer thickness and the bias applied. The authors note that narrowband measurements at 3.4~THz sit at the bottom of an etalon minima, strongly enhancing the modulation effect experienced at that frequency. By carefully choosing the LC layer thickness to control the etalon reflection timings, narrowband operation at a frequency of interest could be enhanced, whereas the use of index-matched polymers to reduce the etalon effects could allow for more broadband devices using this LCD technology.

Whilst power modulation offers immediate potential, liquid crystals that are birefringent at THz frequencies present opportunities for the development of more advanced THz adaptive optical components including variable wave plates, phase shifters and wavefront modulators.
By using a more suitable window material (reduced reflection/absorption at THz-frequencies), anti-reflective coatings, and a liquid crystal material with higher birefringence and linear dichroism (improving both modulation depth and requiring thinner devices, improving modulation speeds), it is expected that significant improvements can be made on these original devices, improving their viability for THz optical components in the future.

\section*{Data availability statement}

Data underlying the results presented in this paper are openly available from the University of Leeds Data Repository~\cite{Dataset}.

\section*{Acknowledgments}
The authors would like to acknowledge financial support from UK Research \& Innovation (Future Leader Fellowship MR/S016929/1), the UK Centre for Earth Observation Instrumentation (Fast Track Contract RP10G0435A03), the UK Space Agency (Pathfinder Contract NSTP3-PF3-078), and the Engineering and Physical Sciences Research Council (EP/P007449/1 and EP/P027687/1).
For the purpose of open access, the author has applied a CC-BY public copyright licence to any Author Accepted Manuscript (AAM) version arising from this submission.

\section*{Author declarations}
\subsection{Conflict of interest}
The authors have no conflicts to disclose.

\subsection{Author contributions}
A.~Dunn: writing --- original draft (lead); data curation (lead); investigation (lead); formal analysis (equal); methodology (equal); validation (equal); visualization (lead). Z.~Zhang: investigation (equal); methodology (equal); resources (equal). M.~D.~Horbury: investigation (supporting); methodology (supporting). E.~V.~Nuttall: investigation (supporting). Y.~Han: investigation (supporting). M.~Salih: investigation (supporting); resources (supporting). L.~Li: investigation (supporting); resources (lead). A.~Bond: conceptualization (supporting); methodology (supporting). E.~Saleh: funding acquisition (supporting); conceptualization (supporting). R.~Harris: funding acquisition (supporting); conceptualization (supporting). N.~Daghestani: funding acquisition (supporting). D.~Pardo: funding acquisition (supporting). B.~N.~Ellison: funding acquisition (supporting). A.~D.~Burnett: investigation (supporting); funding acquisition (supporting); conceptualization (supporting); formal analysis (equal); methodology (equal); validation (equal); supervision (supporting); visualization (supporting); writing --- review and editing (equal). H.~F.~Gleeson: funding acquisition (supporting); conceptualization (supporting); methodology (equal); supervision (supporting); writing --- review and editing (equal). A.~Valavanis: funding acquisition (lead); investigation (supporting); conceptualization (lead); methodology (equal); project administration (lead); supervision (lead); writing --- review and editing (equal).

 
\bibliography{IEEEabrv,main}
%

\bibliographystyle{IEEEtran}











\newpage

 
\vspace{11pt}



\begin{IEEEbiography}[{\includegraphics[width=1in,height=1.25in,clip,keepaspectratio]{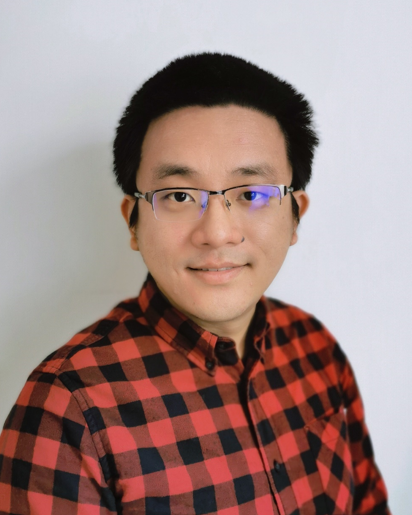}}]{Zhaopeng Zhang}
received the M.Phys. (Hons) degree in physics from the University of Manchester, Manchester. U.K., in 2011 and the Ph.D. degree in physics from the University of Manchester, Manchester. U.K., in 2015.

From 2015--21 he worked as a Research Fellow at the University of Leeds. He is currently a teaching assistant in the School of Electronic and Electrical Engineering. His research experience is in soft-matter materials: liquid crystals, polymers and gels, and applications ranging from photonic devices to soft matter for energy devices.
\end{IEEEbiography}

\begin{IEEEbiography}[{\includegraphics[width=1in,height=1.25in,clip,keepaspectratio]{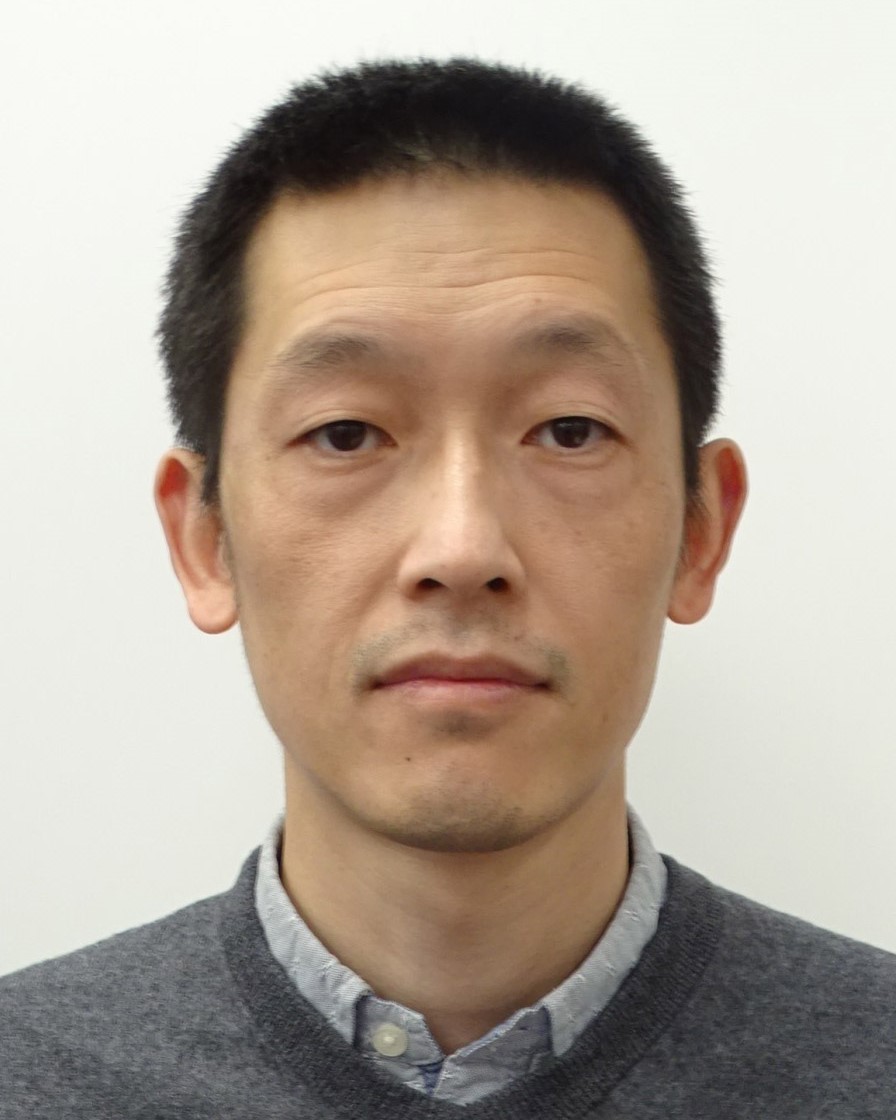}}]{Yingjun Han}
received the B.Eng. degree in automatic control from the Northwestern Polytechnical University, China, in 1998, and the Ph.D. degree in condensed matter physics from the Institute of Physics, Chinese Academy of Sciences, China, in 2003.

He was with the Paul Drude Institute, Germany from 2003--05, the Shanghai Institute of Microsystem and Information Technology, Chinese Academy of Sciences, China, from 2006--12, and the University of Leeds, U.K. from 2012--2021.
He is currently a professor in solid-state terahertz technology at the Shanghai Institute of Microsystem and Information Technology, Chinese Academy of Sciences. His research interests include quantum-cascade lasers, infrared detectors, and terahertz technology.
\end{IEEEbiography}

\begin{IEEEbiographynophoto}{Mohammed Salih}
received the M.Sc. degree in electrical engineering from the University of Kassel, Kassel, Germany, in 2002, and the Ph.D. degree in electronic and electrical engineering from the University of Leeds, Leeds, U.K., in 2011.

Since 2012 he has worked with the University of Leeds, Leeds, U.K., and is currently a Research Fellow. His research interests include the fabrication and characterization of terahertz quantum cascade lasers and terahertz instrumentation for atmospheric and space applications.
\end{IEEEbiographynophoto}

\begin{IEEEbiography}[{\includegraphics[width=1in,height=1.25in,clip,keepaspectratio]{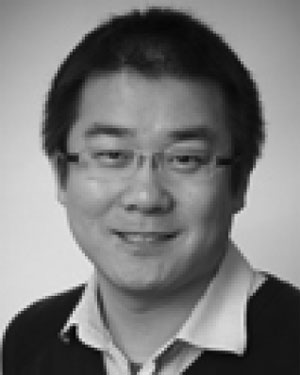}}]{Lianhe Li}
received the Ph.D. degree in microelectronics and solid-state electronics from the Institute of Semiconductors, Chinese Academy of Sciences, Beijing, China, in 2001. 

From 2001 to 2003, he was with the Laboratoire de Photonique et des Nanostructures, Centre National de la Recherche Scientifique (CNRS), France, where he was engaged in molecular beam epitaxy (MBE) growth and characterization of low-bandgap GaAs-based III-V diluted nitride materials and devices with a particular emphasis on 1300–1550 nm telecom wavelength applications. In July 2003, he joined EPFL, Lausanne, Switzerland, as a Scientific Collaborator, where he was working on InAs quantum dots for lasers and superluminescent LEDs and single quantum dot devices. He is currently with the School of Electronic and Electrical Engineering, University of Leeds, U.K 
\end{IEEEbiography}

\begin{IEEEbiography}[{\includegraphics[width=1in,height=1.25in,clip,keepaspectratio]{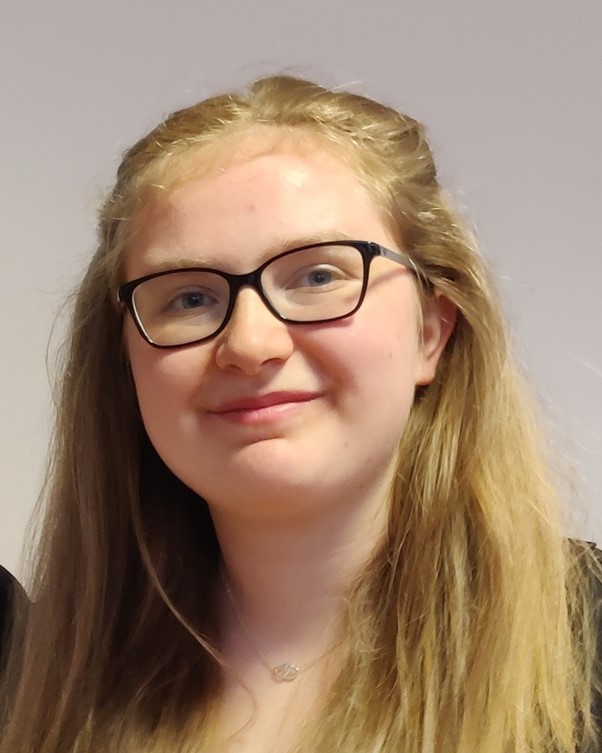}}]{Abigail Bond}
received the M.Phys. (Hons) degree in Physics from the University of Leeds, Leeds, U.K., in 2022.

She is currently a Ph.D. Researcher at the University of Leeds, Leeds, U.K. Her research interests include devices using liquid crystal droplets as chemical sensors.
\end{IEEEbiography}

\begin{IEEEbiography}[{\includegraphics[width=1in,height=1.25in,clip,keepaspectratio]{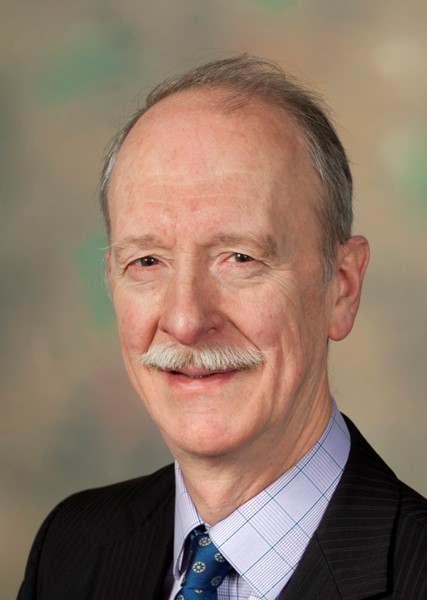}}]{Brian Ellison (FREng, CEng, FIET)}
 received the B.Sc. (Hons) degree in physics and astrophysics from the University of London, in 1977, and an honorary doctorate from The Open University, in 2017.
 
 In 1977 he joined the then UK Science Research Council and worked at its Chilbolton Observatory where he developed microwave radiometers for astronomy research. He moved to the California Institute of Technology, USA, in 1984 and where he developed millimetre-wave superconducting receivers and also acted as Chief Engineer for the Caltech Submillimetre-wave Observatory. In 1989 he returned to the UK where he worked within the Science and Technology Facilities Council (STFC) Rutherford Appleton Laboratory where he eventually led its Millimetre-wave Technology Group until his retirement in 2022. He remains active in the terahertz field and holds visiting scientist roles within STFC and the University of Manchester. He has also recently established CombeTech Ltd, which provides technical consultancy in support of Earth observation and radio astronomy receiver technology.
 
 Professor Ellison was elected a Fellow of the Royal Academy of Engineering in 2019.
 \end{IEEEbiography}

\begin{IEEEbiography}[{\includegraphics[width=1in,height=1.25in,clip,keepaspectratio]{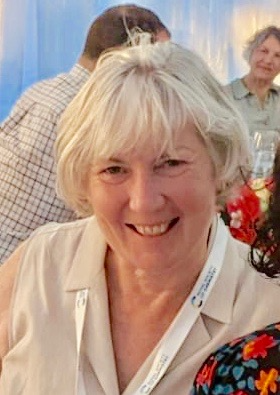}}]{Helen Gleeson}
received the B.Sc. (Hons) degree in Maths and Physics in 1983 and the Ph.D. degree in physics in 1986, both from the University of Manchester.

She became a Lecturer at Manchester in 1989 and a professor in 2003. She has held positions at the University of Manchester, Manchester, U.K., including Dean for Research in Engineering and Physical Science and Head of School of Physics and Astronomy. She took up the Cavendish Chair of Physics at the University of Leeds, Leeds, U.K., in 2015 and was Head of the School of Physics 2016--2021.

Professor Gleeson was awarded the OBE for Services to Science in 2009 and has won awards from the British Liquid Crystal Society and the Institute of Physics. In addition to her research, she is active in matters concerning inclusion and equality in physics. She currently holds an Established Career and ED\&I Champion Fellowship from the EPSRC allowing her to pursue her interests in the physics and applications of liquid crystals.
\end{IEEEbiography}

\begin{IEEEbiography}[{\includegraphics[width=1in,height=1.25in,clip,keepaspectratio]{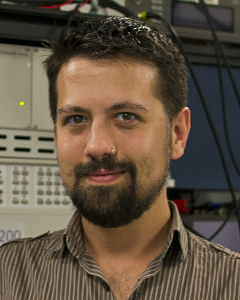}}]{Alexander Valavanis}
received the M.Eng. (Hons) degree in electronic engineering from the University of York, York, U.K., in 2004, and the Ph.D. degree in electronic and electrical engineering from the University of Leeds, Leeds, U.K., in 2009. 

From 2004–5, he was an instrumentation engineer with STFC Daresbury Laboratories, Warrington, U.K., and from 2009–16 he was a Research Fellow at the University of Leeds. He is currently an associate professor in Terahertz Instrumentation at the University of Leeds. His research interests include terahertz instrumentation for atmospheric and space applications, quantum cascade lasers, and computational methods for quantum electronics.

Dr Valavanis is a member of the Institution of Engineering and Technology (IET). 
\end{IEEEbiography}



\vfill

\end{document}